\documentclass[twocolumn,english,aps,pra,showpacs]{revtex4}
\usepackage[T1]{fontenc}
\usepackage[latin1]{inputenc}
\usepackage{amsmath}
\usepackage{amssymb}
\usepackage{graphicx}
\usepackage{esint}

\makeatletter
\@ifundefined{textcolor}{}
{%
 \definecolor{BLACK}{gray}{0}
 \definecolor{WHITE}{gray}{1}
 \definecolor{RED}{rgb}{1,0,0}
 \definecolor{GREEN}{rgb}{0,1,0}
 \definecolor{BLUE}{rgb}{0,0,1}
 \definecolor{CYAN}{cmyk}{1,0,0,0}
 \definecolor{MAGENTA}{cmyk}{0,1,0,0}
 \definecolor{YELLOW}{cmyk}{0,0,1,0}
}

\usepackage{babel}

\usepackage{babel}

\makeatother

\usepackage{babel}
\begin{document}

\title{Momentum-resolved radio-frequency spectroscopy of ultracold atomic
Fermi gases in a spin-orbit coupled lattice}

\author{Xia-Ji Liu$^{1}$}

\affiliation{$^{1}$ARC Centre of Excellence for Quantum-Atom Optics, Centre for
Atom Optics and Ultrafast Spectroscopy, Swinburne University of Technology,
Melbourne 3122, Australia}

\date{\today}
\begin{abstract}
We investigate theoretically momentum-resolved radio-frequency (rf)
spectroscopy of a non-interacting atomic Fermi gas in a spin-orbit
coupled lattice. This lattice configuration has been recently created
at MIT {[}Cheuk \textit{et al.}, arXiv:1205.3483{]} for $^{6}$Li
atoms, by coupling the two hyperfine spin-states with a pair of Raman
laser beams and additional rf coupling. Here, we show that momentum-resolved
rf spectroscopy can measure single-particle energies and eigenstates
and therefore resolve the band structure of the spin-orbit coupled
lattice. In our calculations, we take into account the effects of
temperatures and harmonic traps. Our predictions are to be confronted
with future experiments on spin-orbit coupled Fermi gases of $^{40}$K
atoms in a lattice potential. 
\end{abstract}


\maketitle

\section{Introduction}

The past few years have witnessed an exponential growth of interest
in studying ultracold atomic gases under a synthetic gauge field \cite{Lin2009,Lin2011,Williams2012,JG2012,Chen2012,exptShanXi,exptMIT,Wang2010,Wu2011,Hu2012,Deng2012,Vyasanakere2011,Yu2011,Hu2011,Gong2011,Liu2012}.
The growth is strongly motived by a series of ground-breaking experiments
at the National Institute of Standards and Technology (NIST) \cite{Lin2009,Lin2011,Williams2012,JG2012}.
Most notably, synthetic spin-orbit coupling - the coupling between
the spin and orbital degrees of freedom of the atom - was created
and detected in an atomic Bose-Einstein condensate (BEC) of $^{87}$Rb
atoms in early 2011 \cite{Lin2011}. Such a spin-orbit coupling is
responsible for the recently discovered topological states of matter,
such as topological insulators and spin quantum Hall materials \cite{Qi2010,Hasan2010}
which are new types of functional materials that may lead to novel
quantum devices. It is natural to anticipate that the investigation
of spin-orbit coupled ultracold atomic gases will provide an entirely
new platform to simulate and understand new generation materials.

To date, spin-orbit coupled atomic Fermi gases have been realized
at ShanXi University \cite{exptShanXi} and at the Massachusetts Institute
of Technology (MIT) \cite{exptMIT}, by using fermionic $^{40}$K
atoms and $^{6}$Li atoms, respectively. The technique used to induce
spin-orbit coupling in Fermi gases is more or less the same as in
BECs. A pair of counter-propagating laser beams along $x$-axis is
used to connect two atomic hyperfine spin-states, labeled by $\left|\uparrow\right\rangle $
and $\left|\downarrow\right\rangle $, via a two-photon Raman transition.
The Raman beams impart momentum $2\hbar k_{R}{\bf e}_{x}$ to a fermion
while changing its spin from $\left|\downarrow\right\rangle $ to
$\left|\uparrow\right\rangle $. In this way, the orbital motion is
coupled to spin and an effective spin-orbit coupling is generated.
In the MIT experiment \cite{exptMIT}, an additional radio-frequency
(rf) coupling is used to couple the two hyperfine spin-states. Combined
with the pair of Raman beams, this creates a periodic lattice potential,
in addition to spin-orbit coupling. The rich band structure of such
a novel spin-orbit coupled lattice has been characterized through
spin-injection spectroscopy \cite{exptMIT}, which uses a rf laser
beam to inject free atoms in a third spin state into an empty spin-orbit
coupled system, and then obtains the momentum and spin of injected
atoms using time of flight and spin-resolved detection. The spin-injection
technique is particularly useful for $^{6}$Li atoms. Due to the rapid
heating from Raman process, the spin-orbit coupled Fermi gas of $^{6}$Li
atoms can hardly be created in equilibrium. The heating problem can
be avoided by the spin-injection of atoms from a free Fermi gas in
the third spin state, which do not experience the Raman process. We
note that for $^{40}$K atoms the heating issue due to Raman process
is much milder. As a result, a spin-orbit coupled Fermi gas of $^{40}$K
atoms can be created in equilibrium at $T\simeq0.6T_{F}$ \cite{exptShanXi},
where $T_{F}$ is the Fermi temperature.

In this paper, we investigate theoretically momentum-resolved rf spectroscopy
of a {\em non-interacting}, {\em trapped} atomic Fermi gas of
$^{40}$K atoms in a spin-orbit coupled lattice, given the perspective
that such a system can easily be realized at ShanXi University \cite{exptShanXi}.
This can be viewed as the first step to understand momentum-resolved
rf spectroscopy of a strongly interacting atomic Fermi gas in spin-orbit
coupled lattice. The momentum-resolved rf spectroscopy, whose initial
state is a spin-orbit coupled Fermi gas in equilibrium, yields equivalent
information to spin-injection spectroscopy for a non-interacting system.
However, the latter approach may hardly give useful information for
a strongly-interacting system, due to the lack of equilibrium in the
final spin-orbit coupled state. In our calculations, we take into
account the effect of harmonic traps by using local density approximation.
The effect of temperatures is also addressed.

Our paper is organized as follows. In the next section (Sec. II),
we give the model Hamiltonian of spin-orbit lattice and explain how
to calculate the single-particle energies and eigenstates. In Sec.
III, we derive the expression for momentum-resolved rf-spectroscopy
and discuss the results for a homogeneous spin-orbit coupled system.
In Sec. IV, we present the rf spectroscopy of a trapped system within
local density approximation. We discuss in detail the evolution of
the rf spectroscopy as functions of the temperature, Raman coupling
and rf coupling. Finally, Sec. V is devoted to conclusions and some
final remarks.

\section{Model Hamiltonian}

A non-interacting atomic Fermi gas in a spin-orbit coupled lattice
may be described by the model Hamiltonian \cite{exptMIT}, ${\cal H}={\cal H}_{0}+{\cal H}_{R}+{\cal H}_{RF}$,
where, 
\begin{eqnarray}
{\cal H}_{0} & = & \sum_{\sigma}\int d{\bf r}\psi_{\sigma}^{\dagger}\left({\bf r}\right)\frac{\hbar^{2}k^{2}}{2M}\psi_{\sigma}\left({\bf r}\right),\\
{\cal H}_{R} & = & \frac{\Omega_{R}}{2}\int d{\bf r}\left[\psi_{\uparrow}^{\dagger}\left({\bf r}\right)e^{i2k_{R}x}\psi_{\downarrow}\left({\bf r}\right)+\text{H.c.}\right],\\
{\cal H}_{RF} & = & \frac{\Omega_{RF}}{2}\int d{\bf r}\left[\psi_{\uparrow}^{\dagger}\left({\bf r}\right)\psi_{\downarrow}\left({\bf r}\right)+\text{H.c.}\right].
\end{eqnarray}
 Here, $\psi_{\sigma}^{\dagger}\left({\bf r}\right)$ is the creation
field operator for atoms in the spin-state $\sigma=\left|\uparrow\right\rangle $
and $\left|\downarrow\right\rangle $. The Hamiltonians ${\cal H}_{R}$
and ${\cal H}_{RF}$ describe, respectively, the pair of counter-propagating
Raman laser beams and the additional rf coupling that couple the two
hyperfine spin-states. $\Omega_{R}$ is the Raman coupling strength,
$k_{R}$ $=2\pi/\lambda$ is determined by the wave length $\lambda$
of two lasers and $2\hbar k_{R}$ is the momentum transfer during
the two-photon Raman process, $\Omega_{RF}$ is the rf coupling strength.

The Hamiltonian ${\cal H}_{R}$ creates the spin-orbit coupling, and
with ${\cal H}_{RF}$, a spin-orbit coupled system in a lattice potential
can be formed. To see this, let us take the following gauge transformation,
\begin{eqnarray}
\psi_{\uparrow}\left({\bf r}\right) & = & e^{+ik_{R}x}\tilde{\psi}_{\uparrow}\left({\bf r}\right),\\
\psi_{\downarrow}\left({\bf r}\right) & = & e^{-ik_{R}x}\tilde{\psi}_{\downarrow}\left({\bf r}\right),
\end{eqnarray}
 with which the model Hamiltonians become, 
\begin{eqnarray}
{\cal H}_{0} & = & \sum_{\sigma}\int d{\bf r}\left[\tilde{\psi}_{\sigma}^{\dagger}\left({\bf r}\right)\frac{\hbar^{2}\left({\bf k}\pm k_{R}{\bf e}_{x}\right)^{2}}{2M}\tilde{\psi}_{\sigma}\left({\bf r}\right)\right],\\
{\cal H}_{R} & = & \frac{\Omega_{R}}{2}\int d{\bf r}\left[\tilde{\psi}_{\uparrow}^{\dagger}\left({\bf r}\right)\tilde{\psi}_{\downarrow}\left({\bf r}\right)+\text{H.c.}\right],\\
{\cal H}_{RF} & = & \frac{\Omega_{RF}}{2}\int d{\bf r}\left[\tilde{\psi}_{\uparrow}^{\dagger}\left({\bf r}\right)e^{-i2k_{R}x}\tilde{\psi}_{\downarrow}\left({\bf r}\right)+\text{H.c.}\right],
\end{eqnarray}
 where in the first term of ${\cal H}_{0}$ we take ``$+$'' for
spin-up atoms and ``$-$'' for spin-down atoms. By introducing a
spinor field operator $\Phi({\bf r})\equiv[\tilde{\psi}_{\uparrow}\left({\bf r}\right),\tilde{\psi}_{\downarrow}\left({\bf r}\right)]^{T}$
and using the Pauli matrices $\sigma_{x}$, $\sigma_{y}$, and $\sigma_{z}$,
we can write compactly the model Hamiltonian in the form, 
\begin{equation}
{\cal H}=\int d{\bf r}\Phi^{\dagger}\left({\bf r}\right)\left[H_{SO}+V_{L}\left(x\right)\right]\Phi\left({\bf r}\right),
\end{equation}
 where we have defined the spin-orbit Hamiltonian 
\begin{equation}
H_{SO}\equiv\frac{\hbar^{2}\left(k_{R}^{2}+{\bf k}^{2}\right)}{2M}+h\sigma_{x}+\lambda k_{x}\sigma_{z}
\end{equation}
 and the rf lattice potential 
\begin{equation}
V_{L}\left(x\right)\equiv V_{L}\left[\cos\left(2k_{R}x\right)\sigma_{x}+\sin\left(2k_{R}x\right)\sigma_{y}\right].
\end{equation}
 Here, for convenience we have introduced a spin-orbit coupling constant
$\lambda\equiv\hbar^{2}k_{R}/M$, an ``effective'' Zeeman field
$h\equiv\Omega_{R}/2$, and an ``effective'' lattice depth $V_{L}\equiv\Omega_{RF}/2$.

\subsection{Single-particle solution for $H_{SO}$}

The model Hamiltonian $H_{SO}$ describes a spin-orbit coupling with
equal Rashba and Dresselhaus strengths \cite{Lin2011,Chen2012,exptShanXi,exptMIT}.
The single-particle solution $\phi_{{\bf k}}({\bf r)}$ satisfies
the Schrödinger equation, $H_{SO}\phi_{{\bf k}}({\bf r)=}\epsilon_{{\bf k}}\phi_{{\bf k}}({\bf r)}$.
Using the Pauli matrices and the fact that the wave-vector or momentum
${\bf k}\equiv(k_{x},{\bf k}_{\perp})\equiv(k_{x},k_{y},k_{z})$ is
a good quantum number, it is easy to see that we have two eigenvalues
\begin{equation}
\epsilon_{{\bf k\pm}}=\frac{\hbar^{2}k_{\perp}^{2}}{2M}+\frac{\hbar^{2}\left(k_{R}^{2}+k_{x}^{2}\right)}{2M}{\bf \pm}\sqrt{h^{2}+\lambda^{2}k_{x}^{2}},\label{soEnergy}
\end{equation}
 where ``${\bf \pm}$'' stands for two helicity branches. The corresponding
eigenstates are given by (we set the volume $V=1$), 
\begin{eqnarray}
\phi_{{\bf k}}^{\left(+\right)}\left({\bf r}\right) & = & \left[\left(\begin{array}{c}
\cos\theta_{{\bf k}}\\
\sin\theta_{{\bf k}}
\end{array}\right)e^{ik_{x}x}\right]e^{i{\bf k}_{\perp}\cdot{\bf r}_{\perp}},\\
\phi_{{\bf k}}^{(-)}\left({\bf r}\right) & = & \left[\left(\begin{array}{c}
-\sin\theta_{{\bf k}}\\
\cos\theta_{{\bf k}}
\end{array}\right)e^{ik_{x}x}\right]e^{i{\bf k}_{\perp}\cdot{\bf r}_{\perp}},
\end{eqnarray}
 where $\theta_{{\bf k}}=\arctan[(\sqrt{h^{2}+\lambda^{2}k_{x}^{2}}-\lambda k_{x})/h]$
and ${\bf r}_{\perp}\equiv(y,z)$.

\subsection{Single-particle solution for the spin-orbit coupled lattice}

In the presence of the additional rf Hamiltonian ${\cal H}_{RF}$,
the momentum along the $x$-axis, $k_{x}$, is no longer a good quantum
number. The lattice potential terms $\cos\left(2k_{R}x\right)$ and
$\sin\left(2k_{R}x\right)$ will couple the eigenstates $\phi_{{\bf k}^{\prime}}^{(\pm)}\left({\bf r}\right)$
and $\phi_{{\bf k}^{\prime\prime}}^{(\pm)}\left({\bf r}\right)$ if
$k_{x}^{\prime}-k_{x}^{\prime\prime}=2nk_{R}$, where $n=\pm1,\pm2,\cdots$
is an integer. In this case, it is useful to define a quasi-momentum
or lattice momentum $q_{x}$ for arbitrary $k_{x}$ as follows: $k_{x}=2nk_{R}+q_{x}$,
where the integer $n$ is chosen to make $-k_{R}\leq q_{x}<k_{R}$.
The quasi-momentum $q_{x}$ is then a good quantum number. We may
expand the single-particle eigenstate of the total Hamiltonian in
the form, 
\begin{equation}
\Phi\left(q_{x},{\bf k}_{\perp};{\bf r}\right)=\sum_{m=-\infty}^{+\infty}\left[a_{m+}\phi_{{\bf k}_{m}}^{(+)}\left({\bf r}\right)+a_{m-}\phi_{{\bf k}_{m}}^{(-)}\left({\bf r}\right)\right],\label{wfexpansion}
\end{equation}
 where ${\bf k}_{m}\equiv{\bf k}_{\perp}+(2mk_{R}+q_{x}){\bf e}_{x}\equiv{\bf k}_{\perp}+k_{mx}{\bf e}_{x}$
has the same quasi-momentum $q_{x}$, and the energies of $\phi_{{\bf k}_{m}}^{(+)}\left({\bf r}\right)$
and $\phi_{{\bf k}_{m}}^{(-)}\left({\bf r}\right)$ are given by 
\begin{equation}
\epsilon_{m{\bf \pm}}\equiv\frac{\hbar^{2}k_{\perp}^{2}}{2M}+\frac{\hbar^{2}\left(k_{R}^{2}+k_{mx}^{2}\right)}{2M}{\bf \pm}\sqrt{h^{2}+\lambda^{2}k_{mx}^{2}}.
\end{equation}
 The coefficients $a_{n+}$ and $a_{n-}$ can be determined by the
Schrödinger equation, 
\begin{equation}
\left[H_{SO}+V_{L}\left(x\right)\right]\Phi\left(q_{x},{\bf k}_{\perp};{\bf r}\right)=E(q_{x},{\bf k}_{\perp})\Phi\left(q_{x},{\bf k}_{\perp};{\bf r}\right),
\end{equation}
 where $E(q_{x},{\bf k}_{\perp})\equiv E(q_{x})+\hbar^{2}k_{\perp}^{2}/(2M)$.
By substituting the wave-function (\ref{wfexpansion}) into the above
Schrödinger equation and multiplying on both sides $\phi_{{\bf k}_{n}}^{(\pm)}\left({\bf r}\right)$
and finally taking the integration $\int d{\bf r}$, it is straightforward
to show that,\begin{widetext} 
\begin{equation}
\left(\begin{array}{cc}
\epsilon_{n{\bf +}} & 0\\
0 & \epsilon_{n{\bf -}}
\end{array}\right)\left(\begin{array}{l}
a_{n+}\\
a_{n-}
\end{array}\right)+\sum_{m=-\infty}^{+\infty}\left(\begin{array}{cc}
V_{nm}^{++} & V_{nm}^{+-}\\
V_{nm}^{-+} & V_{nm}^{--}
\end{array}\right)\left(\begin{array}{l}
a_{m+}\\
a_{m-}
\end{array}\right)=\left[\frac{\hbar^{2}k_{\perp}^{2}}{2M}+E\left(q_{x}\right)\right]\left(\begin{array}{l}
a_{n+}\\
a_{n-}
\end{array}\right),\label{secularEQ}
\end{equation}
 where the matrix elements of the rf Hamiltonian are given by, 
\begin{eqnarray}
V_{nm}^{++} & = & V_{L}\left(+\delta_{n+1,m}\cos\theta_{{\bf k}_{n}}\sin\theta_{{\bf k}_{m}}+\delta_{n,m+1}\sin\theta_{{\bf k}_{n}}\cos\theta_{{\bf k}_{m}}\right),\\
V_{nm}^{+-} & = & V_{L}\left(+\delta_{n+1,m}\cos\theta_{{\bf k}_{n}}\cos\theta_{{\bf k}_{m}}-\delta_{n,m+1}\sin\theta_{{\bf k}_{n}}\sin\theta_{{\bf k}_{m}}\right),\\
V_{nm}^{-+} & = & V_{L}\left(-\delta_{n+1,m}\sin\theta_{{\bf k}_{n}}\sin\theta_{{\bf k}_{m}}+\delta_{n,m+1}\cos\theta_{{\bf k}_{n}}\cos\theta_{{\bf k}_{m}}\right),\\
V_{nm}^{--} & = & V_{L}\left(-\delta_{n+1,m}\sin\theta_{{\bf k}_{n}}\cos\theta_{{\bf k}_{m}}-\delta_{n,m+1}\cos\theta_{{\bf k}_{n}}\sin\theta_{{\bf k}_{m}}\right).
\end{eqnarray}
\end{widetext} For a given quasi-momentum $q_{x}$, therefore the
wave-functions $a_{n+}$ and $a_{n-}$ and the corresponding energy
$E\left(q_{x}\right)$ can be obtained from the secular equation (\ref{secularEQ})
by exact diagonalization. In the numerical calculation, we have to
make a truncation, i.e., let $-N_{\max}\leq n,m\leq N_{\max}$, where
$N_{\max}$ is a large integer. We will take $k_{R}$ and $E_{R}\equiv\hbar^{2}k_{R}^{2}/(2M)$
as the units for (quasi)-momentum and energy, respectively.

\begin{figure}[htp]
\begin{centering}
\includegraphics[clip,width=0.45\textwidth]{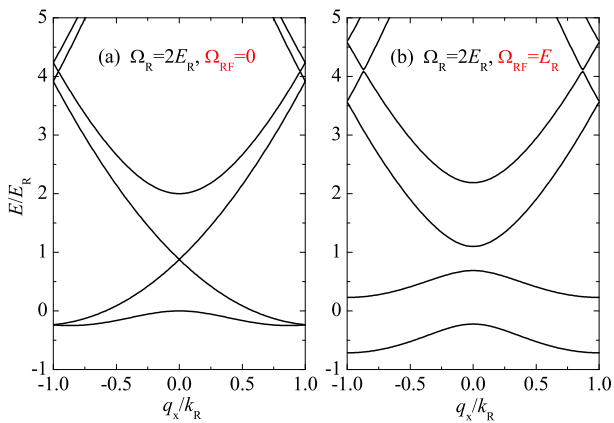} 
\par\end{centering}

\caption{(color online) Energy band structure $E\left(q_{x}\right)$ in the
absence (a) and presence (b) of spin-orbit lattice potential.}

\label{fig1} 
\end{figure}

\subsection{Band structure of spin-orbit coupled lattice}

In Fig. 1, we present the band structure $E\left(q_{x}\right)$ of
the spin-orbit coupled system at $\Omega_{RF}=0$ and $\Omega_{RF}=E_{R}$.
In the absence of rf-coupling (Fig. 1a), the band structure is actually
exactly identical to the single-particle dispersion Eq. (\ref{soEnergy}).
However, we have folded the entire dispersion into the first Brillouin
zone $-k_{R}\leq q_{x}<k_{R}$. Thus, at the edge of Brillouin zone,
$q_{x}=\pm k_{R}$, the band energy is at least two-fold degenerate.
Moreover, some bands are also degenerate at $q_{x}=0$ due to the
even-parity of the single-particle dispersion (\ref{soEnergy}). In
this case, for the wave function Eq. (\ref{wfexpansion}), there is
only one non-zero (i.e., unity) coefficient in $a_{n+}$ or $a_{n-}$.
With the rf-coupling (Fig. 1b), the degeneracy at the zone edge or
at $q_{x}=0$ is lifted. We have a clear, well-resolved band structure.
When the rf-coupling $\Omega_{RF}$ is large enough, a band gap also
opens among the lowest three bands. Therefore, when the system is
filled up to the top of the second band (i.e., the chemical potential
$\mu<E_{R}$), a band insulator is formed.

\section{Radio-frequency spectroscopy in free space}

Let us consider the rf-spectroscopy \cite{Chin2004,Schunck2008},
which is driven by a rf laser beam to transfer an atom in one of the
two hyperfine states (say $\left|\downarrow\right\rangle $) to an
empty hyperfine state $\left|3\right\rangle $. The state $\left|3\right\rangle $
is normally higher in energy by an amount of $\hbar\omega_{3\downarrow}$,
due to the magnetic field splitting in bare atomic hyperfine levels.
The Hamiltonian for the rf-transition may be written as, 
\begin{eqnarray}
{\cal V}_{rf} & = & V_{0}\int d{\bf r}\left[\psi_{3}^{\dagger}\left({\bf r}\right)\psi_{\downarrow}\left({\bf r}\right)+\psi_{\downarrow}^{\dagger}\left({\bf r}\right)\psi_{3}\left({\bf r}\right)\right],\\
 & = & V_{0}\int d{\bf r}\left[e^{-ik_{R}x}\psi_{3}^{\dagger}\left({\bf r}\right)\tilde{\psi}_{\downarrow}\left({\bf r}\right)+\text{H.c.}\right],
\end{eqnarray}
 where $\psi_{3}^{\dagger}\left({\bf r}\right)$ is the field operator
which creates an atom in $\left|3\right\rangle $ at the position
${\bf r}$ and $V_{0}$ is the strength of the rf drive. In the second
line of the above equation, we have taken the gauge transformation
for $\psi_{\downarrow}\left({\bf r}\right)$. As a result, there is
an effective momentum transfer $k_{R}{\bf e}_{x}$.

The transfer strength of the rf-transition $\Gamma\left(\omega\right)$
can be calculated by using the Fermi's golden rule:\begin{widetext}
\begin{equation}
\Gamma\left(\omega\right)=\sum_{i,f}\left|\left\langle \Phi_{f}\right|{\cal V}_{rf}\left|\Phi_{i}\right\rangle \right|^{2}f\left(E_{i}-\mu\right)\delta\left[\hbar\omega-\hbar\omega_{3\downarrow}-\left(E_{f}-E_{i}\right)\right].
\end{equation}
 Here, the summation is over all the possible initial states $\Phi_{i}$
(with energy $E_{i}$) and final states $\Phi_{f}$ (with energy $E_{f}$)
and $f\left(E_{i}-\mu\right)$ is the Fermi distribution function.
The Dirac delta function ensures energy conservation during transition.
Hereafter, without any confusion we shall ignore the energy splitting
in the bare atomic hyperfine levels and set $\omega_{3\downarrow}=0$.
To calculate the overlap between the initial and final wave-functions
$\left|\left\langle \Phi_{f}\right|{\cal V}_{rf}\left|\Phi_{i}\right\rangle \right|$,
let us take the $l$-th band eigenstate $\Phi_{i}=\Phi^{(l)}\left(q_{x},{\bf k}_{\perp};{\bf r}\right)\equiv[\tilde{\psi}_{\uparrow}^{(l)}\left({\bf r}\right),\tilde{\psi}_{\downarrow}^{(l)}\left({\bf r}\right)]^{T}$
as the initial state, where 
\begin{equation}
\tilde{\psi}_{\downarrow}^{(l)}\left({\bf r}\right)=\sum_{n=-\infty}^{+\infty}\left[a_{n+}^{(l)}\sin\theta_{{\bf k}_{n}}+a_{n-}^{(l)}\cos\theta_{{\bf k}_{n}}\right]e^{i\left(2nk_{R}+q_{x}\right)x}e^{i{\bf k}_{\perp}\cdot{\bf r}_{\perp}},
\end{equation}
 and $E_{i}=E^{(l)}(q_{x})+\hbar^{2}k_{\perp}^{2}/(2M)$. It is easy
to see that, in order to have a nonzero overlap, the final state must
be a plane wave, i.e., $e^{i({\bf k}_{n}-k_{R}{\bf e}_{x})\cdot{\bf r}}$,
with which the overlap of wave functions is given by, 
\begin{equation}
\left|\left\langle \Phi_{f}\right|{\cal V}_{rf}\left|\Phi_{i}\right\rangle \right|^{2}=\left[a_{n+}^{(l)}\sin\theta_{{\bf k}_{n}}+a_{n-}^{(l)}\cos\theta_{{\bf k}_{n}}\right]^{2},
\end{equation}
 and the final state energy is 
\begin{equation}
E_{f}=\frac{\hbar^{2}k_{\perp}^{2}}{2M}+\frac{\hbar^{2}\left(k_{nx}-k_{R}\right)^{2}}{2M}.
\end{equation}
 By taking into account all the possibilities for $\Phi_{i}$ and
$\Phi_{f}$, the transfer strength can then be written in the form,
\begin{eqnarray}
\Gamma\left(\omega\right) & = & \int\limits _{0}^{\infty}\frac{k_{\perp}dk_{\perp}}{\left(2\pi\right)^{2}}\int\limits _{-k_{R}}^{+k_{R}}dq_{x}\sum_{l=0}^{\infty}\sum_{n=-\infty}^{+\infty}\left[a_{n+}^{(l)}\sin\theta_{{\bf k}_{n}}+a_{n-}^{(l)}\cos\theta_{{\bf k}_{n}}\right]^{2}\nonumber \\
 &  & \times f\left[\frac{\hbar^{2}k_{\perp}^{2}}{2M}+E^{(l)}(q_{x})-\mu\right]\delta\left[\hbar\omega+E^{(l)}(q_{x})-\frac{\hbar^{2}\left(k_{nx}-k_{R}\right)^{2}}{2M}\right],
\end{eqnarray}
 where $k_{nx}=2nk_{R}+q_{x}$ and $\theta_{{\bf k}_{n}}=\arctan[(\sqrt{h^{2}+\lambda^{2}k_{nx}^{2}}-\lambda k_{nx})/h]$.
The integration over $k_{\perp}$ can be done analytically. We find
that, 
\begin{equation}
\int\limits _{0}^{\infty}\frac{k_{\perp}dk_{\perp}}{\left(2\pi\right)^{2}}f\left[\frac{\hbar^{2}k_{\perp}^{2}}{2M}+E^{(l)}(q_{x})-\mu\right]=\frac{Mk_{B}T}{4\pi^{2}\hbar^{2}}\ln\left\{ 1+\exp\left[-\frac{E^{(l)}(q_{x})-\mu}{k_{B}T}\right]\right\} .
\end{equation}
 Experimentally, the momentum of the transferred atom, $k_{nx}$,
could be resolved \cite{exptMIT,Stewart2008}. Therefore, we may define
a momentum-resolved transfer strength ($k_{x}\equiv k_{nx}-k_{R}$),
\begin{equation}
\Gamma\left(k_{x},\omega\right)=\frac{Mk_{B}T}{4\pi^{2}\hbar^{2}}\sum_{l=0}^{\infty}\left[a_{n+}^{(l)}\sin\theta_{{\bf k}_{n}}+a_{n-}^{(l)}\cos\theta_{{\bf k}_{n}}\right]^{2}\ln\left\{ 1+\exp\left[-\frac{E^{(l)}(q_{x})-\mu}{k_{B}T}\right]\right\} \delta\left[\hbar\omega+E^{(l)}(q_{x})-\frac{\hbar^{2}k_{x}^{2}}{2M}\right],\label{homoRF}
\end{equation}
\end{widetext} where the quasi-momentum $q_{x}$ and the index $n$
are now determined from the momentum $k_{nx}=k_{x}+k_{R}$. The total
transfer strength is simply $\Gamma(\omega)=\int\nolimits _{-\infty}^{+\infty}dk_{x}\Gamma(k_{x},\omega)$.
To take into account the energy resolution of the spectroscopy $\gamma\sim0.1E_{R}$
\cite{exptMIT}, we may replace the Dirac delta function by, $\delta\left(x\right)=(\gamma/\pi)/[x^{2}+\gamma^{2}]$.
To reveal more clearly the band structure, it is also useful to calculate
\begin{equation}
\tilde{\Gamma}\left(k_{nx},\tilde{\omega}\right)\equiv\Gamma\left(k_{x}+k_{R},\omega+\frac{\hbar k_{x}^{2}}{2M}\right),
\end{equation}
 for which, the Dirac delta function takes the form $\delta[\hbar\tilde{\omega}+E^{(l)}(q_{x})]$.

\subsection{Momentum-resolved rf spectroscopy in free space}

We are now ready to calculate the momentum-resolved rf spectroscopy
of a uniform Fermi gas in the spin-orbit coupled lattice. For a given
$k_{x}$, we obtain first the quasi-momentum $q_{x}$ and the index
$n$, and then solve the energy bands $E^{(l)}(q_{x})$ and eigenstates
$a_{n+}^{(l)}$ and $a_{n-}^{(l)}$. For the given chemical potential
$\mu$ and temperature $k_{B}T$, which are in units of $E_{R}$,
we finally take the summation over the band index $l$ and obtain
the momentum-resolved transfer strength. Figs. 2a-2c and 2d-2f report
the linear contour plot of rf spectroscopy in the absence ($\Omega_{RF}=0$)
and presence ($\Omega_{RF}=E_{R}$) of lattice potential, respectively,
with increasing the chemical potential $\mu$ at zero temperature.
The transfer strength always becomes stronger with increasing $\mu$,
since there are more and more atoms in the system.

\begin{figure}[htp]
\begin{centering}
\includegraphics[clip,width=0.45\textwidth]{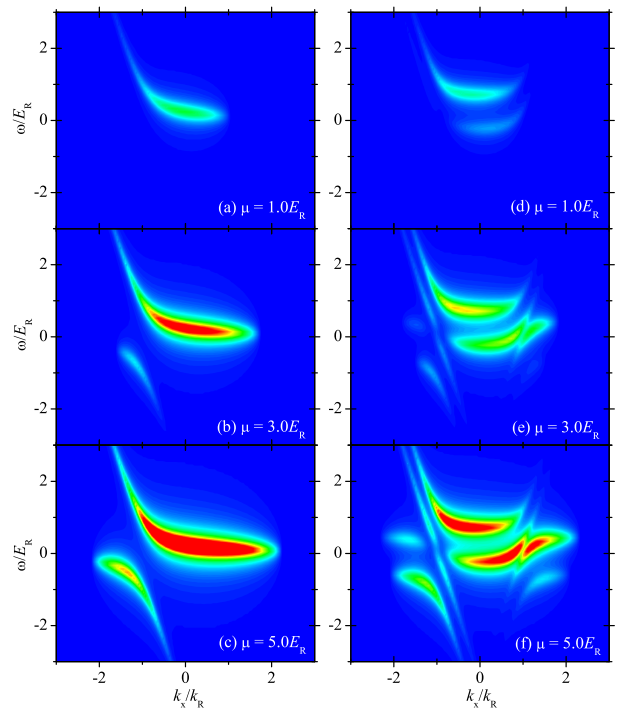} 
\par\end{centering}

\caption{(color online) Momentum-resolved rf spectroscopy of a uniform Fermi
gas without (left panel, $\Omega_{RF}=0$) or with (right panel, $\Omega_{RF}=E_{R}$)
lattice potential at zero temperature. The Raman coupling strength
is $\Omega_{R}=2E_{R}$. From (a) to (c), or from (d) to (f), we increase
the chemical potential $\mu$ from $E_{R}$ to $5E_{R}$. Here the
intensity of the contour plot increases linearly from $0$ (blue)
to $0.1M/\hbar^{2}$ (red).}

\label{fig2} 
\end{figure}

The spectroscopy is already very interesting without a lattice, as
shown in Figs. 2a-2c. At a low chemical potential $\mu=E_{R}$, only
the lower helicity branch of the single-particle dispersion Eq. (\ref{soEnergy})
is occupied. The spin-orbit coupling leads to a long tail at negative
momentum and high frequency, in sharp contrast to a single Dirac delta
function $\delta(\omega)$ in the absence of spin-orbit coupling.
As the chemical potential increases (Figs. 2b and 2c), the upper helicity
branch gets occupied. In the spectroscopy, this creates a strong response
at opposite momentum and frequency.

In the presence of a lattice potential induced by the rf-coupling,
the spectroscopy is greatly modified by the formation of energy bands.
At the chemical potential $\mu=E_{R}$ the lowest two bands, well-separated
in energy, should already be occupied, as we can see from Fig. 1b.
As a result, we observe in Fig. 2d the two different responses from
the two bands. The energy gap between the first and second bands,
which is at about $E_{R}$, is clearly resolved in the spectroscopy.
With increasing the chemical potential, more and more energy bands
come to contribute and the spectroscopy becomes more fragmented. There
are some discontinuity at $k_{x}=\pm k_{R}$, indicating the existence
of different energy gaps.

\begin{figure}[htp]
\begin{centering}
\includegraphics[clip,width=0.45\textwidth]{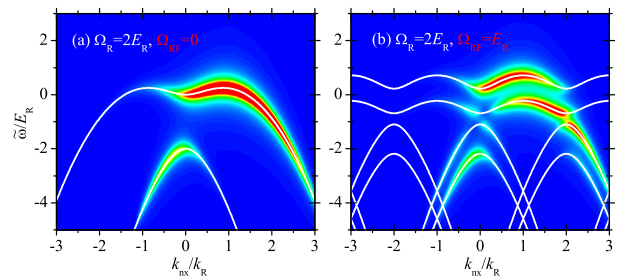} 
\par\end{centering}

\caption{(color online) Linear contour plot of $\tilde{\Gamma}(k_{nx},\tilde{\omega})$
at $\Omega_{R}=2E_{R}$, $\mu=5E_{R}$, and $T=0$. The left plot
and right plot correspond to the cases without and with lattice potential,
respectively. The energy bands $-E^{(l)}(q_{x})$ are shown by thick
white curves.}

\label{fig3} 
\end{figure}

Fig. 3 shows $\tilde{\Gamma}(k_{nx},\tilde{\omega})$ at $\mu=5E_{R}$
with or without the lattice potential. As $\tilde{\Gamma}(k_{nx},\tilde{\omega})$
contains the Dirac delta function $\delta[\hbar\tilde{\omega}+E^{(l)}(q_{x})]$,
we anticipate that the energy band $E^{(l)}(q_{x})$ can be seen clearly
from the contour plot. Indeed, we find that the rf-response is peaked
exactly at $-E^{(l)}(q_{x})$ (shown by white curves) within the experimental
energy resolution. The strength of the response is determined by the
coefficients $a_{n+}^{(l)}$ and $a_{n-}^{(l)}$. Therefore, by measuring
momentum-resolved rf-spectroscopy $\Gamma\left(k_{x},\omega\right)$
and re-constructing $\tilde{\Gamma}(k_{nx},\tilde{\omega})$, we are
able to obtain the complete information of the single-particle energy
bands and eigenstates.

\section{Radio-frequency spectroscopy in harmonic traps}

We turn to address the realistic issue of harmonic traps, $V_{T}({\bf r})=M\omega_{0}^{2}r^{2}/2$,
by using local density approximation. Within local density approximation,
the whole system can be regarded as a collection of many uniform blocks
with a local chemical potential $\mu-V_{T}({\bf r})$. The momentum-resolved
rf spectroscopy is a sum of local spectroscopy over the whole trap,
\begin{equation}
\Gamma_{T}\left(k_{x},\omega\right)=\int d{\bf r}\Gamma\left[k_{x},\omega;\mu-V_{T}({\bf r})\right].
\end{equation}
 By substituting Eq. (\ref{homoRF}) for local spectroscopy and using
the fact that\begin{widetext} 
\begin{equation}
\int_{0}^{\infty}4\pi r^{2}dr\ln\left\{ 1+\exp\left[-\frac{E^{(l)}(q_{x})-\mu+V_{T}({\bf r})}{k_{B}T}\right]\right\} =-\left(\frac{2\pi k_{B}T}{M\omega_{0}^{2}}\right)^{3/2}\text{Li}_{5/2}\left(-\exp\left[-\frac{E^{(l)}(q_{x})-\mu}{k_{B}T}\right]\right),
\end{equation}
 we find that, 
\begin{eqnarray}
\Gamma_{T}\left(k_{x},\omega\right) & = & -\sqrt{\frac{E_{R}}{\pi}}\frac{\left(k_{B}T\right)^{5/2}}{\left(\hbar\omega_{0}\right)^{3}}\sum_{l=0}^{\infty}\left[a_{n+}^{(l)}\sin\theta_{{\bf k}_{n}}+a_{n-}^{(l)}\cos\theta_{{\bf k}_{n}}\right]^{2}\nonumber \\
 &  & \times\text{Li}_{5/2}\left(-\exp\left[-\frac{E^{(l)}(q_{x})-\mu}{k_{B}T}\right]\right)\delta\left[\hbar\omega+E^{(l)}(q_{x})-\frac{\hbar^{2}k_{x}^{2}}{2M}\right].
\end{eqnarray}
\end{widetext} Here Li$_{n}(x)$ is the polylogarithm function. Experimentally,
the trap frequency is about two order smaller than the recoil energy
$E_{R}$. Hereafter, we shall take $\hbar\omega_{0}=0.01E_{R}$, according
to the experimental setup at ShanXi University \cite{exptShanXi}.
In analogy to the uniform case, we may define 
\begin{equation}
\tilde{\Gamma}_{T}\left(k_{nx},\tilde{\omega}\right)\equiv\Gamma_{T}\left(k_{x}+k_{R},\omega+\frac{\hbar k_{x}^{2}}{2M}\right),
\end{equation}
 in order to better visualize the energy band.

\begin{figure}[htp]
\begin{centering}
\includegraphics[clip,width=0.45\textwidth]{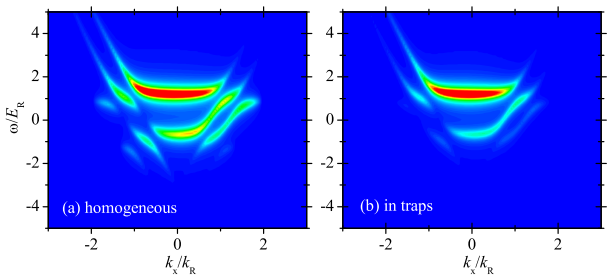} 
\par\end{centering}

\caption{(color online) Comparison between momentum-resolved rf-spectroscopy
in free space and in harmonic traps at zero temperature. Here, we
take the same chemical potential $\mu=3E_{R}$ and use $\Omega_{R}=$
$\Omega_{RF}=2E_{R}$. The intensity of each contour plot increases
from $0$ (blue) to its maximum value (red).}

\label{fig4} 
\end{figure}

In Figs. 4a and 4b, we compare the momentum-resolved rf spectroscopy
in free space and in harmonic traps, at $\Omega_{R}=$ $\Omega_{RF}=2E_{R}$
and at zero temperature. The chemical potential is taken the same
value, i.e., $\mu=3E_{R}$. In each plot, we assign the red color
to the maximum value of rf transfer strength. With traps, the rf-response
from the higher bands is blurred by the trap average. However, the
qualitative features of rf-spectroscopy remains the same, as we may
anticipate. This strongly indicates that in harmonic traps we could
still be able to re-construct the energy band structure by using momentum-resolved
rf-spectroscopy.

\begin{figure}[htp]
\begin{centering}
\includegraphics[clip,width=0.45\textwidth]{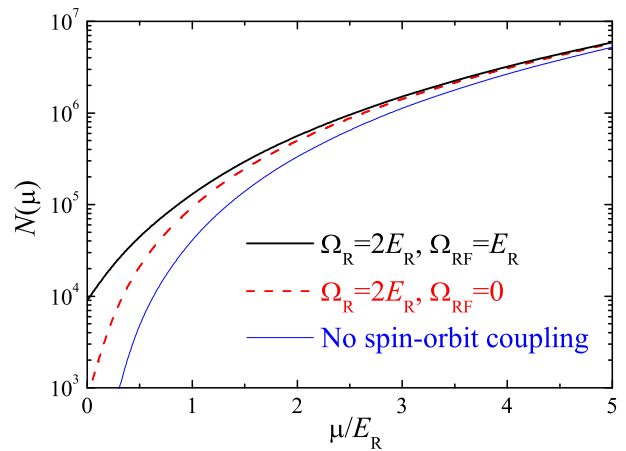} 
\par\end{centering}

\caption{(color online) Total number of atoms as a function of chemical potential
for a trapped Fermi gas. The cases with or without lattice potential
are plotted by solid and dashed lines, respectively. For comparison,
we show also the result without spin-orbit coupling by a thin solid
line. Here, we take the trapping frequency $\hbar\omega_{0}=0.01E_{R}$.}

\label{fig5} 
\end{figure}

To have a realistic estimate of the chemical potential in harmonic
traps, it is useful to calculate the total number of atoms, which
is given by $N=\int d{\bf r}n({\bf r})$, where the local density\begin{widetext}
\begin{equation}
n({\bf r})=\int\limits _{0}^{\infty}\frac{k_{\perp}dk_{\perp}}{\left(2\pi\right)^{2}}\int\limits _{-k_{R}}^{+k_{R}}dq_{x}\sum_{l=0}^{\infty}f\left[\frac{\hbar^{2}k_{\perp}^{2}}{2M}+E^{(l)}(q_{x})-\mu+V_{T}({\bf r})\right].
\end{equation}
 By integrating over $k_{\perp}$ and the spatial coordinates, we
obtain, 
\begin{equation}
N=-\sqrt{\frac{E_{R}}{\pi}}\frac{\left(k_{B}T\right)^{5/2}}{\left(\hbar\omega_{0}\right)^{3}}\int\limits _{-k_{R}}^{+k_{R}}dq_{x}\sum_{l=0}^{\infty}\text{Li}_{5/2}\left(-\exp\left[-\frac{E^{(l)}(q_{x})-\mu}{k_{B}T}\right]\right).
\end{equation}
\end{widetext} In Fig. 5, we plot the total number of atoms in harmonic
traps as a function of chemical potential with or without the spin-orbit
coupled lattice. For comparison, we show also the result for an ideal
Fermi gas without any spin-orbit coupling, $N=\mu^{3}/[3(\hbar\omega_{0})^{3}]$.
In the ShanXi experiment \cite{exptShanXi}, the number of atoms is
about $2\times10^{6}$, corresponding to $\mu\sim3E_{R}$.

\subsection{Momentum-resolved rf spectroscopy in harmonic traps}

\begin{figure}[htp]
\begin{centering}
\includegraphics[clip,width=0.4\textwidth]{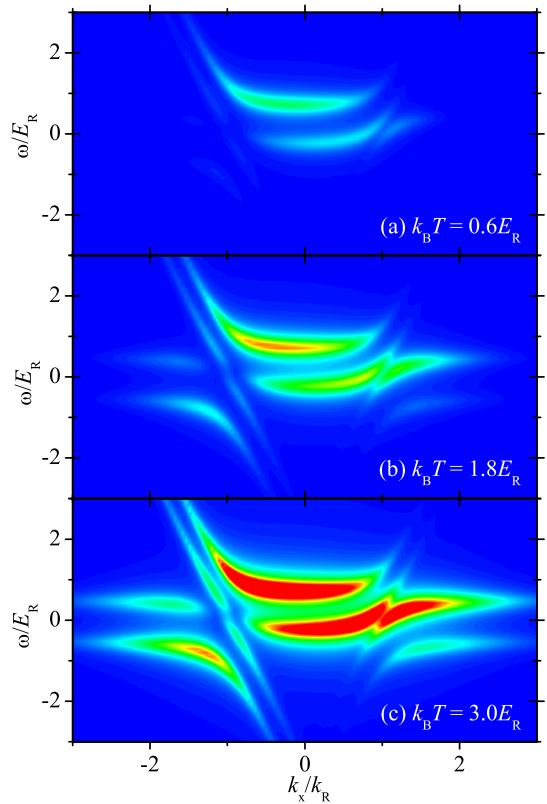} 
\par\end{centering}

\caption{(color online) Temperature dependence of the momentum-resolved rf
spectroscopy of a trapped atomic Fermi gas at $\Omega_{R}=2E_{R}$,
$\Omega_{RF}=E_{R}$ and $\mu=3E_{R}$. The intensity of the contour
plots increases from 0 (blue) to $2.5E_{R}^{3}/(\hbar\omega_{0})^{3}$
(red).}

\label{fig6} 
\end{figure}

We examine first how the rf-spectroscopy is affected by temperature.
In Fig. 6, we report the evolution of rf-spectroscopy with increasing
temperature $k_{B}T$ from $0.2\mu$, $0.6\mu$ to $\mu$, where $\mu=3E_{R}$,
$\Omega_{R}=2E_{R}$, and $\Omega_{RF}=E_{R}$. As the temperature
increases, more and more energy bands are visible, as these bands
become thermally occupied. Importantly, there is no significant thermal
broadening for higher energy bands. They are all well-resolved even
close to the degeneracy temperature $k_{B}T\sim\mu$. Note that, the
typical temperature in the ShanXi experiment is about $0.6k_{B}T_{F}\sim0.6\mu$.

\begin{figure*}[tph]
\begin{centering}
\includegraphics[clip,width=0.95\textwidth]{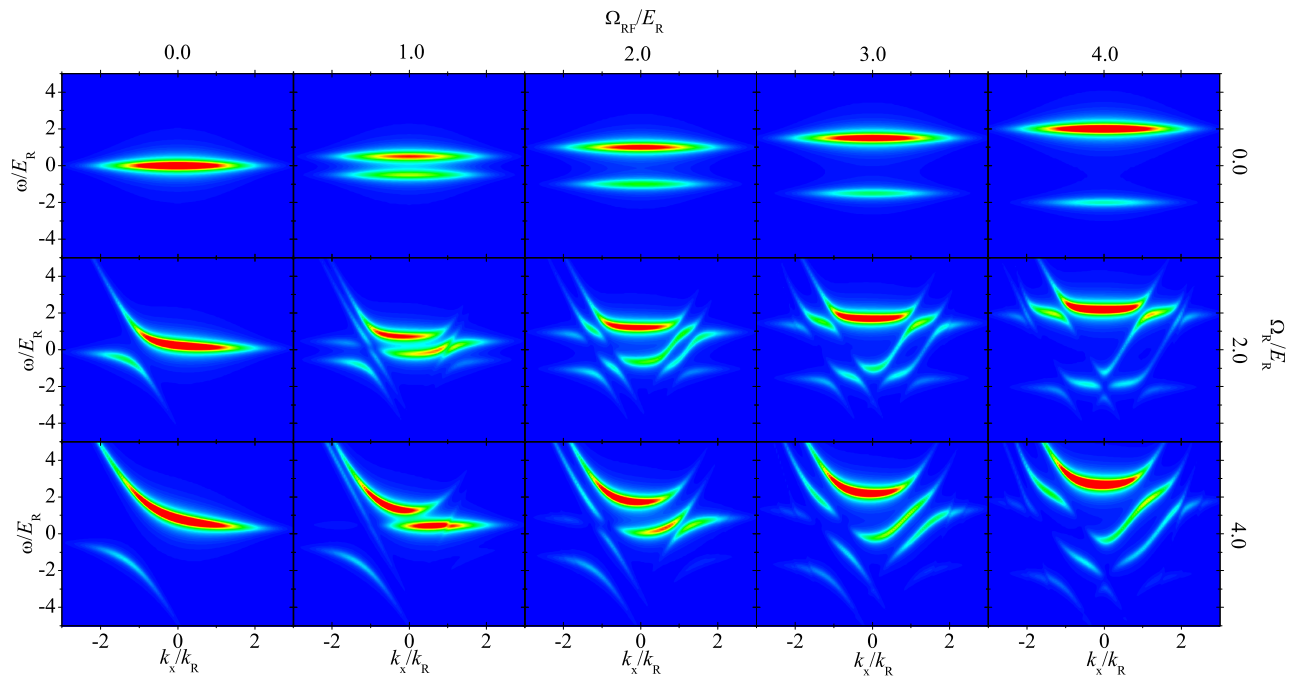} 
\par\end{centering}

\caption{(color online) Evolution of the momentum-resolved rf spectroscopy
of a trapped atomic Fermi gas as functions of the Raman and rf coupling
strengths. Here, we take $\mu=3E_{R}$ and $k_{B}T=0.6\mu$. The intensity
of the contour plots increases from 0 (blue) to $2E_{R}^{3}/(\hbar\omega_{0})^{3}$
(red).}

\label{fig7} 
\end{figure*}

We now explore the spin-orbit coupled system for a range of coupling
strengths. In Fig. 7, we report the evolution of rf-spectroscopy as
functions of the Raman coupling $\Omega_{R}$ and the rf coupling
$\Omega_{RF}$, at $k_{B}T=0.6\mu$ and $\mu=3E_{R}$. This may be
viewed as a realistic simulation of a future experiment for a non-interacting
trapped Fermi gas of $^{40}$K atoms in spin-orbit coupled lattice
\cite{exptShanXi}. 

The reconstructed plot of $\tilde{\Gamma}_{T}(k_{nx},\tilde{\omega})$
is shown in Fig. 8, for a set of parameters $\Omega_{R}=\Omega_{RF}=2E_{R}$,
$\mu=3E_{R}$, and $T=0.6\mu$. By comparing with the energy bands
$-E^{(l)}(q_{x})$, which is plotted by thick white curves, it is
readily seen that the band structure can be clearly extracted from
a realistic momentum-resolved measurement at finite temperatures (i.e.,
$T\sim0.6T_{F}$) and finite energy resolutions (i.e., $\gamma\sim0.1E_{R}$). 

\begin{figure}[htp]
\begin{centering}
\includegraphics[clip,width=0.45\textwidth]{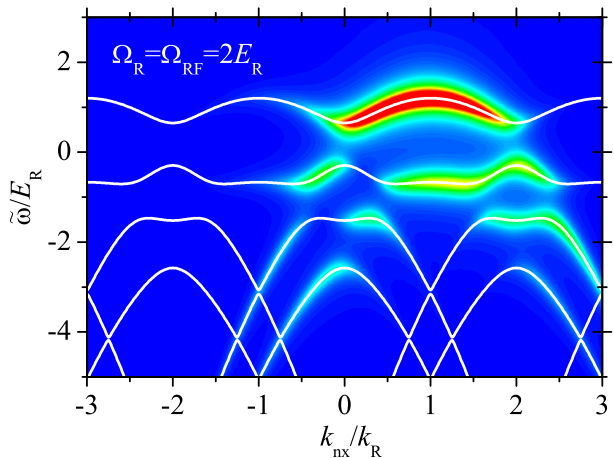} 
\par\end{centering}

\caption{(color online) Linear contour plot of $\tilde{\Gamma}_{T}(k_{nx},\tilde{\omega})$
at $\Omega_{R}=\Omega_{RF}=2E_{R}$, $\mu=3E_{R}$, and $T=0.6\mu$.
The energy bands $-E^{(l)}(q_{x})$ are shown by thick white curves.
The intensity of the contour plots increases from 0 (blue) to $1.5E_{R}^{3}/(\hbar\omega_{0})^{3}$
(red).}

\label{fig8} 
\end{figure}

\section{Conclusions}

In summary, we have predicted theoretically momentum-resolved rf spectroscopy
of a non-interacting atomic Fermi gas in a spin-orbit coupled lattice.
We have shown that such a rf-spectroscopy, just like the angle-resolved
photoemission spectroscopy in condensed matter physics \cite{ARPES},
provides an ideal technique to characterize the non-trivial band structure
of spin-orbit coupled lattice. Our predictions can be readily examined
at ShanXi University by using ultracold $^{40}$K atoms \cite{exptShanXi}.

Using Feshbach resonances \cite{Chin2010}, a strongly interacting
Fermi gas of $^{40}$K atoms in spin-orbit coupled lattice would be
created experimentally very soon. It is of great interest to study
how the single-particle band structure is modified by strong interatomic
interactions and fermionic superfluidity. In this case, we anticipate
that momentum-resolved rf spectroscopy would provide very useful information.

\section*{Acknowledgments}

We thank very much Hui Hu for useful discussions and for helps on
preparing the figures. This work is supported by the ARC Discovery
Project (Grant No. DP0984637) and the NFRP-China (Grant No. 2011CB921502).

\end{document}